\documentclass[12pt]{article}
\usepackage{amsmath,amssymb,amsfonts,color,graphicx,cite,feynarts,soul}
\usepackage[usenames,dvipsnames,svgnames,table]{xcolor}
\input paperdef

\graphicspath{{figs/}}

\evensidemargin \oddsidemargin
\allowdisplaybreaks

\begin{document}
\thispagestyle{empty}

\def\thefootnote{\fnsymbol{footnote}}

\begin{flushright}
TUM--HEP--937/14
\mbox{}
\end{flushright}

\vspace{0.5cm}

\begin{center}

{\large\sc {\bf Does the LHC exclude SUSY Particles at the ILC?
}}%
\footnote{Talk presented by S.H.\ at the International Workshop on
  Future Linear Colliders (LCWS13), Tokyo, Japan, 11-15 November 2013.} 
\vspace{1cm}

{\sc
A.~Bharucha$^{1}$%
\footnote{email: Aoife.Bharucha@tum.de}%
, S.~Heinemeyer$^{2}$%
\footnote{email: Sven.Heinemeyer@cern.ch}%
, F.~von der Pahlen$^{2,3}$%
\footnote{email: fp@gfif.udea.edu.co}%
\footnote{MultiDark Fellow}%
}

\vspace*{.7cm}

{\sl
$^1$ Physik Department T31, Technische Universit\"at M\"unchen, James-Franck-Stra\ss e~1, D-85748 Garching, Germany
\vspace{0.1cm}

$^2$Instituto de F\'isica de Cantabria (CSIC-UC), E-39005 Santander, Spain
\vspace{0.1cm}

$^3$Instituto de F\'isica, Universidad de Antioquia, Calle 70 No.~52-21, Medell\'in, Colombia

}
\end{center}

\vspace*{0.1cm}

\begin{abstract}
\centerline{No.}

\end{abstract}
\vspace*{.5cm}

\def\thefootnote{\arabic{footnote}}
\setcounter{page}{0}
\setcounter{footnote}{0}

\newcommand{\DecCCh[1]}{\cha{2} \to \cha{1} h_{#1}}
\newcommand{\DecCCZ}{\cha{2} \to \cha{1} Z}
\newcommand{\DecCNH}[2]{\cha{#1} \to \neu{#2} H^\mp}
\newcommand{\DecCNW}[2]{\cha{#1} \to \neu{#2} W^\mp}
\newcommand{\DecCnSl}[3]{\cham{#1} \to \nu_{#2}\, \tilde{#2}_{#3}^{-}}
\newcommand{\DecClSn}[2]{\cham{#1} \to {#2}^{-}\, \tilde{\nu_{#2}}}
\newcommand{\DecCxy}[1]{\cha{#1} \to {\rm xy}}

\newcommand{\DecNCW}[2]{\neu{#1} \to \champ{#2} W^\pm}
\newcommand{\DecNCH}[2]{\neu{#1} \to \champ{#2} H^\pm}
\newcommand{\DecNlpmSl}[3]{\neu{#1} \to {#2}^\mp\, \tilde{#2}_{#3}^\pm}
\newcommand{\DecNnSnnbarSn}[2]{\neu{#1} \to \bar{\nu}_{#2}\, \tilde{\nu}_{#2}  /  {\nu}_{#2}\, \tilde{\nu}_{#2}^\dagger  }

\newcommand{\decCCh}{\DecCCh{k}}
\newcommand{\decCCZ}{\DecCCZ}
\newcommand{\decCNH}{\DecCNH{i}{j}}
\newcommand{\decCNW}{\DecCNW{i}{j}}
\newcommand{\decCnSl}{\DecCnSl{i}{l}{k}}
\newcommand{\decClSn}{\DecClSn{i}{l}}
\newcommand{\decCxy}{\DecCxy{i}}

\newcommand{\DecNxy}[1]{\neu{#1} \to {\rm xy}}
\newcommand{\DecNCmH}[2]{\neu{#1} \to \cham{#2} H^+}
\newcommand{\DecNCmW}[2]{\neu{#1} \to \cham{#2} W^+}
\newcommand{\DecNCpH}[2]{\neu{#1} \to \chap{#2} H^-}
\newcommand{\DecNCpW}[2]{\neu{#1} \to \chap{#2} W^-}
\newcommand{\DecNNh}[3]{\neu{#1} \to \neu{#2} h_{#3}}
\newcommand{\DecNNZ}[2]{\neu{#1} \to \neu{#2} Z}
\newcommand{\DecNnSn}[2]{\neu{#1} \to {\nu}_{#2}\, \tilde{\nu}_{#2}^\dagger}
\newcommand{\DecNlSl}[3]{\neu{#1} \to {#2}^-\, \tilde{#2}_{#3}^+}
\newcommand{\DecNbarnSn}[2]{\neu{#1} \to \bar{\nu}_{#2}\, \tilde{\nu}_{#2}}
\newcommand{\DecNlpSl}[3]{\neu{#1} \to {#2}^+\, \tilde{#2}_{#3}^-}

\newcommand{\decNCH}{\DecNCH{i}{j}}
\newcommand{\decNCW}{\DecNCW{i}{j}}
\newcommand{\decNlpmSl}{\DecNlpmSl{i}{\ell}{k}}
\newcommand{\decNnSnnbarSn}{\DecNnSnnbarSn{i}{\ell}}

\newcommand{\decNCmH}{\DecNCmH{i}{j}}
\newcommand{\decNCmW}{\DecNCmW{i}{j}}
\newcommand{\decNNh}{\DecNNh{i}{j}{k}}
\newcommand{\decNNZ}{\DecNNZ{i}{j}}
\newcommand{\decNnSn}{\DecNnSn{i}{\ell}}
\newcommand{\decNlSl}{\DecNlSl{i}{\ell}{k}}
\newcommand{\decNbarnSn}{\DecNbarnSn{i}{\ell}}
\newcommand{\decNlpSl}{\DecNlpSl{i}{\ell}{k}}

\newcommand{\decNxy}{\DecNxy{i}}

\newcommand{\DecCmCh}[1]{\cham{2} \to \cham{1} h_{#1}}
\newcommand{\DecCmCZ}{\cham{2} \to \cham{1} Z}
\newcommand{\DecCmNH}[2]{\cham{#1} \to \neu{#2} H^-}
\newcommand{\DecCmNW}[2]{\cham{#1} \to \neu{#2} W^-}
\newcommand{\DecCmnSl}[3]{\cham{#1} \to \bar{\nu}_{#2}\, \tilde{#2}_{#3}^-}
\newcommand{\DecCmlSn}[2]{\cham{#1} \to {#2}^-\, \tilde{\nu}_{#2}^\dagger}
\newcommand{\DecCmxy}[1]{\cham{#1} \to {\rm xy}}

\newcommand{\decCmCh}{\DecCmCh{k}}
\newcommand{\decCmCZ}{\DecCmCZ}
\newcommand{\decCmNH}{\DecCmNH{i}{j}}
\newcommand{\decCmNW}{\DecCmNW{i}{j}}
\newcommand{\decCmnSl}{\DecCmnSl{i}{l}{k}}
\newcommand{\decCmlSn}{\DecCmlSn{i}{l}}
\newcommand{\decCmxy}{\DecCmxy{i}}

\newpage

\section{Introduction}

The LHC is actively searching for physics beyond the Standard Model
(BSM), where the Minimal Supersymmetric Standard Model
(MSSM)~\cite{mssm} is one of the leading candidates.
The search for supersymmetry (SUSY) at the LHC has not (yet) led to
a positive result. 
In particular, bounds on the first and second generation squarks and the gluinos
from ATLAS and CMS are very roughly at the TeV scale, depending on
details of the assumed parameters, see e.g.~\cite{non-deg-squarks}. 
On the other hand, bounds on the
electroweak SUSY sector, where $\cha{1,2}$ and $\neu{1,2,3,4}$ denote
the charginos and neutralinos (i.e.\ the charged and neutral SUSY
partners of the SM gauge and Higgs bosons) are substantially
weaker. 

There are several good motivations to expect electroweak SUSY particles
with masses around a few hundred GeV.
\begin{itemize}
\item
Models based on Grand Unified Theories (GUTs) naturally predict a
lighter electroweak 
spectrum (see \citere{guts} and references therein). 

\item
The anomalous magnetic moment of the muon shows a more than $\sim 4\si,$
deviation from the SM prediction, see \citere{fredl-gm2} and references
therein. Agreement of this measurement with the MSSM requires
charginos and neutralinos in the range of several hundreds of~GeV.

\item
Charginos and neutralinos in the few hundred~GeV range (possibly
together with not too heavy scalar leptons) could easily bring the
prediction of the $W$~boson mass in full agreement with experimental
data, see \citere{MW-lisa} and references therein.

\item
On the more speculative side, light charginos and neutralinos could also
explain the small discrepancies in the $pp \to W^+W^- + X$ measurements
visible in all published data from 
ATLAS and CMS, see \citere{WW-Rolbiecki} and references therein.

\end{itemize}

These points provide a strong motivation for the search of these electroweak
particles, which could be in the kinematic reach of the LHC.
ATLAS~\cite{ATLAS:2012ku,ATLAS:2012-154,ATLAS:2013-028,ATLAS:2013-035,ATLAS:2013-049,Aad:2014vma} 
and CMS~\cite{CMS:2012ewa,CMS:2012ewb} are actively searching for the direct
production of charginos and neutralinos, in particular for the process
$pp \to \cha{1}\neu{2}$ with the subsequent decays 
$\cha{1} \to \neu{1} W^\pm$ and $\neu{2} \to \neu{1} Z$, 
resulting a three lepton signature.
These searches are performed mostly in so-called ``simplified models'',
where the branching ratios of the relevant SUSY particles are set
to one, assuming that all other potential decay modes are kinematically
forbidden.

Based on those (and similar) analyses strong claims about excluded mass
regions, e.g.\ in the $\mneu2$--$\mneu1$ plane are made. This kind of
bounds apparently exclude to a large extent the production of light
charginos and/or neutralinos at the $e^+e^-$ International Linear
Collider (ILC), which is expected to operate with a center-of-mass
energy of $\sqrt{s} \le 1 \tev$.
On the other hand, 
if the charginos and neutralinos were to lie within reach of the ILC, 
it should be possible to reconstruct some or all of the
fundamental parameters describing the electroweak sector of the MSSM
(see e.g.~\cite{Bharucha:2012ya}). 
Here we briefly review LHC exclusion bounds on electroweak particles, in
particular in the $\cha1 \neu2$ search, and their ``correct''
interpretation as exclusions bounds~\cite{Bharucha:2013epa}.%
\footnote{
We will not discuss LHC bounds on colored particles, which often appear to be
strong, but where it is crucial to keep in mind the assumptions made to obtain
these bounds.}


\section{What does the LHC (really) exclude?}

In the case that the coloured sector is heavy, the direct production of
charginos and neutralinos might provide 
the largest cross-sections of SUSY particles at the LHC. 
The golden channel for SUSY production is of chargino neutralino
($\cha{1}\neu{2}$) pair production, dominated by the
$s$-channel $W$~boson mode.
In the absence of light sleptons, 
the  $\cha{1}$ decays via $W^+\neu{1}$, while the $\neu{2}$ may decay
either to a Higgs or a $Z$~boson, and a $\neu{1}$.

However, the experimental results for these decay channels are usually
interpreted in terms of a simplified model which assumes 
100\% branching franction to the $Z$, an assumption which leads to an 
incorrect interpretation once the decay to the Higgs boson is open.
Here we review the analysis of the effect on the exclusion limits when
the branching ratio to the Higgs is included. 
We will first define our notation and discuss the details of the calculation, 
then discuss simplified expressions for the couplings relevant for the
$\neu{2}$ decays, and finally present some results for the reinterpreted
limits, considering both the dependence on $\tb$ (the ratio of the two
vacuuum expectation values of the two MSSM Higgs doublets) and the phase
of the $U(1)$ gaugino mass parameter, $\phiMe$. 


\subsection{Details of the calculation}
\label{sec:notation}

The parameters entering the chargino--neutralino sector are the bino
mass $M_1$, the wino mass $M_2$, the higgsino mass $\mu$ as well as
$\tb$, and $\cw = \MW/\MZ$ and $\sw = \sqrt{1 - \cw^2}$ where $\MZ$ is
the mass of the $Z$~boson. 
When working in the complex MSSM the parameters $M_1$, $M_2$
and $\mu$ can in principle have a non-zero complex phase.
However, one of the phases of these parameters, here $\phiMz$,  can be
rotated away, in which case 
the phase $\phimu$ is tightly constrained~\cite{plehnix}. Therefore we 
take $\mu$ to be a real parameter.
Further note that in the case of the complex MSSM, the three neutral Higgs
bosons $h$, $H$ and $A$ mix at the loop
level~\cite{mhiggsCPXgen,Demir,mhiggsCPXRG1,mhiggsCPXFD1},
resulting in the (mass ordered) $\He$, $\Hz$ and $\Hd$,
which are not states of definite~$\cp$-parity.
In the following we denote the light Higgs with $\He$, independent
whether the parameters are chosen complex or real. The Higgs sector
predictions have been derived with 
{\tt FeynHiggs~2.9.4}~\cite{feynhiggs,mhiggslong,mhiggsAEC,mhcMSSMlong}
(the most recent corrections to the Higgs boson masses as derived in
\citere{Mh-logresum} are not included, but expect to have a small impact
on the parameters used here, see below).

\medskip
In the limit
$\mu\gg |\MOne|, \MTwo$; $\MHp\gg\MZ$; $\tb\gg 1$, 
which will be relevant for most of the analyzed benchmark scenarios, we
obtained simplified expressions for the couplings for
$\neu{i}\neu{j}Z/\He$ in Ref.~\cite{Bharucha:2013epa}.  
Here the two lightest neutralinos are almost
purely bino and wino-like states, 
 $\neu{1}\sim\tilde{B}$, $\neu{2}\sim\tilde{W}$. 
For simplicity we neglect the mixing between the bino and wino components,
which has a subleading effect in our approximation, such that 
$N_{12} \simeq N_{21}\simeq 0$, while $|N_{11}| \simeq |N_{22}| \simeq 1$. 
Note that in the Higgs decoupling limit~\cite{decoupling} 
one has $(\be - \al) \to \pi/2$. 
In this limit we obtain
\begin{align}
C^L_{\neu{1}\neu{2}Z} 
&\approx 
\frac{e}{2} \frac{\MZ^2 }{\mu^2}\exp\left(\frac{i\phiMe}{2}\right) 
~,
\label{eq:nnz.approx}
\\
C^L_{\neu{1}\neu{2}h_1}
&\approx
\frac{e}{2}\frac{\MZ}{\mu} 
\KL \frac{\MOne + \MTwo}{\mu} + \frac{4}{\tb} \KR
\exp\left(\frac{-i\phiMe}{2}\right) 
~, 
\label{eq:nnh.approx}
\end{align}
where the neglected terms are of higher order in $\MZ/\mu,\ M_{1/2}/\mu$
and $ 1/\tb$. Here $e$ denotes the electric charge, 
$\al_{\rm em} = e^2/(4\,\pi)$, and
$\al$ is the angle that diagonalizes the
$\cp$-even Higgs sector at tree-level. 
From \refeq{eq:nnh.approx} it follows
that the absolute value of the Higgs coupling is largest (smallest) for
positive (negative) $\MOne$. 
Note that the partial decay widths, for which explicit expressions
can be found in \citere{Bharucha:2013epa}, also depend on the relative
intrinsic $\cp$ of the neutralinos, and that of the Higgs or
($\cp$-even) $Z$-boson.
Near the decay threshold this effect, which arises due to the p-wave
suppression of some of the amplitudes, leads to a stronger dependence on
the $\cp$-phases than the one resulting from the change in the absolute
value of the couplings, provided $\mneu1 \neq 0$, as shown in
\citere{Bharucha:2013epa}.

Before reviewing our results, we will first briefly describe the
calculations used for the direct production cross section of $\neu{2}\cha{1}$, 
and for the branching ratios for the subsequent decay of the neutralino
into a $Z$ boson and of the chargino into a $W$ boson and the LSP. 
The production of neutralinos and charginos at
the LHC is calculated using the program {\tt Prospino~2.1}~\cite{prospinoNN}. 
Complex parameters can only affect these cross sections when the charginos or
neutralinos are mixed states, and we estimate such effects to be
negligible for our set-up, so we adopt the {\tt Prospino} results which
neglect CP phases to hold. We further neglect the NLL corrections to the
gaugino production cross section calculated in
\citere{Fuks:2012qx,resummino}, estimating their effects to be at the
per-cent level. 
For the decay widths we employ our full NLO corrections in the on-shell
scheme of the complex MSSM (see e.g.~\cite{aoife-decays}) as calculated
in \citeres{LHCxC,LHCxN}. The calculation is based on 
{\tt FeynArts}/{\tt Formcalc}~\cite{feynarts,formcalc}, and the corresponding
model file conventions~\cite{feynarts-mf} are used throughout.
The benchmark scenarios defined in the
following section are such that the decays 
$\cha1 \to \neu1 W^\pm$ as well as $\neu2 \to \neu1 Z$, 
$\neu2 \to \neu1 \He$ are the only relevant ones.
As analyzed in the previous subsection
the decays of a wino-like $\neu{2}$ to $\neu{1} h_i$ 
are most sensitive to $\phiMe$ due to the relative $\cp$ between the
bino-like $\neu{1}$ 
and the wino-like $\neu2$, which is controlled by $\phiMe$.


\subsection{Definition of benchmark scenarios}

The baseline analysis in \citere{Bharucha:2013epa} made use of results
reported by ATLAS using the full 2012 data set, where numerical values
were provided for the excluded cross sections~\cite{ATLAS:2013-035} (but
could equally be applied on more recent Analyses, such as in
\citere{Aad:2014vma}). 
In order to interpret the ATLAS exclusions in terms of the complex MSSM,
we calculate the cross section in benchmark scenarios similar to those
used by ATLAS, including NLO corrections as described below.
We re-analyze the ATLAS
$95\%$~CL exclusion bounds in the simplified analyses in the
$\mneu{2}$--$\mneu{1}$ plane, taking $\MOne$ and $\MTwo$ as free
parameters with central values:
\begin{align}
 M_1=100 \gev \; \mbox{and} \; M_2=250 \gev.
\label{ATLpara1}
\end{align} 
The other parameters are chosen as in the ATLAS analysis
presented in \citere{ATLAS:2013-035}%
\footnote{
Not all parameters are clearly defined in \citere{ATLAS:2013-035}. We
select and choose our parameters to be as close to the original analysis
as possible.}%
, 
\begin{align}
\mu = 1 \tev, \; \tb = 6, \; \Msqez = \Msqd = \Msl = 2 \tev, \; 
\At = 2.8 \tev~.
\label{ATLpara}
\end{align}
$\Msqez$ denotes the diagonal soft SUSY-breaking parameter in the scalar
quark mass matrices of the first and second generation, similarly
$\Msqd$ for the third generation and $\Msl$ for all three generations of
scalar leptons. If all three mass scales are identical we also use the
abbreviation $\msusy := \Msqez = \Msqd = \Msl$. 
$\At$ is the 
trilinear coupling between stop quarks and Higgs bosons, which is
chosen to give the desired value of $\MHe$.
The other trilinear couplings, set to zero in
\citere{ATLAS:2012-154,ATLAS:2013-035}, 
we set to $\At$ for squarks and to zero for sleptons.
Setting also the $A_{q\neq t}$ to zero would have a minor impact on
our analysis.
The effect the large sfermion mass scale is a small destructive
interference of the $s$-channel amplitude with the $t$-channel squark
exchange. 
The large higgsino mass parameter $\mu$ results in a gaugino-like pair of
produced neutralino and chargino. The lightest Higgs boson mass (as
calculated with 
{\tt FeynHiggs~2.9.4}~\cite{feynhiggs,mhiggslong,mhiggsAEC,mhcMSSMlong})
is evaluated to be $\sim 125.5 \gev$, defining the value of $\At$ in
\refeq{ATLpara}. 
In order to scan the $\mneu{2}$--$\mneu{1}$ plane we 
use the ranges
\begin{align}
|\MOne| = 0 \ldots 200 \gev~, \quad \MTwo = 100 \ldots 400 \gev
\quad \mbox{with~} |\MOne| \le \MTwo~.
\end{align}

\medskip
The main aim of \citere{Bharucha:2013epa}, as discussed above, was the
interpretation of 
the ATLAS exclusion limits in several ``physics motivated'' benchmark
scenarios. Taking the parameters in \refeq{ATLpara} as our baseline
scenario, deviations are made in the following directions.

\begin{enumerate}

\item We take $\phiMe$, the phase of $\MOne$, to be a free
 parameter. Note that for the considered central benchmark 
 scenario, as $\tb$ is low and $\msusy$ is high, 
 the full range is allowed by current electric dipole moment (EDM) 
constraints~\cite{Baker:2006ts,Regan:2002ta,Griffith:2009zz},
as verified explicitly via both
 \texttt{CPsuperH~2.3}~\cite{Lee:2012wa,Lee:2007gn,Lee:2003nta} and  
 \texttt{FeynHiggs~2.9.4}~\cite{feynhiggs,mhiggslong,mhiggsAEC,mhcMSSMlong}.

\item The variation of $\tb$ can have a strong impact on the
 couplings between the neutralinos and the Higgs boson,
see \refeq{eq:nnh.approx}. 
We therefore analyze the effect of variation of $\tb$ in the range
$\tb = 6 \ldots 20$.

\end{enumerate}

The various scenarios are summarized in \refta{tab:para}

\newcommand{\Satlas}{$S_{\rm ATLAS}$}
\newcommand{\Scompl}{$S_{\rm ATLAS}^{\phiMe}$}
\newcommand{\nSgauge}{$S_{\tilde{H}\tilde{B}}$}
\newcommand{\Stb}{$S_{\rm ATLAS}^{\tan\be}$}
\newcommand{\Stbt}{$S_{\rm ATLAS}^{\tan\be=20}$}
\newcommand{\Ssusy}{$S_{\rm ATLAS}^{\rm SUSY}$}
\newcommand{\Sstau}{$S^{\rm DM}$}
\newcommand{\Satlasmu}{$S_{\rm ATLAS}^{\mu}$}

\newcommand{\oScompl}{$S_{\rm complex}$}
\newcommand{\Sgauge}{$S_{{\rm low-}\mu}$}
\newcommand{\oStb}{$S_{\tan\be}$}
\newcommand{\oSsusy}{$S_{\rm SUSY}$}
\newcommand{\oSstau}{$S_{\rm DM}$}

\begin{table}[ht!]
\renewcommand{\arraystretch}{1.5}
\BC
\begin{tabular}{|c||c|c|c|c|c|c|
|c|
}
\hline
Scenario & $\phiMe$ & $\mu$& $\tb$ & $\msusy$ 
\\ \hline\hline
\Satlas & 
$ 0$          & $ 1000            $ & $ 6 $ 
& $ 2000 $               %
\\ \hline\hline
\Scompl  &  
$ 0 \ldots \pi$ & $ 1000            $ & $ 6 $ 
& $ 2000 $              %
\\ \hline
\Stb  &  
$ 0              $ & $ 1000             $ & $ 6 \ldots 20$ 
& $ 2000 $              %
\\ \hline
\end{tabular}
\caption{
Parameters for benchmark scenarios (masses in $\gev$). 
We furthermore have for all scenarios: $|\MOne| = 0 \ldots 200 \gev$,
$\MTwo = 100 \ldots 400 \gev$ (with $|\MOne| \le \MTwo$), 
$M_3 = 1500 \gev$ (gluino mass parameter).
The first (baseline)
scenario corresponds to the ATLAS analysis in \citere{ATLAS:2013-035}.
Our ``central benchmark scenario'' refers to the case $\MOne=100 \gev$ and
$\MTwo=250 \gev$. The value of $\At$ is adjusted
to ensure $\MHe \approx 125.5 \gev$.
}
\label{tab:para}
\EC
\renewcommand{\arraystretch}{1.0}
\end{table}


\subsection{Results}\label{sec:res}

\begin{figure}[t!]
\begin{center}
\includegraphics[width=0.70\textwidth]{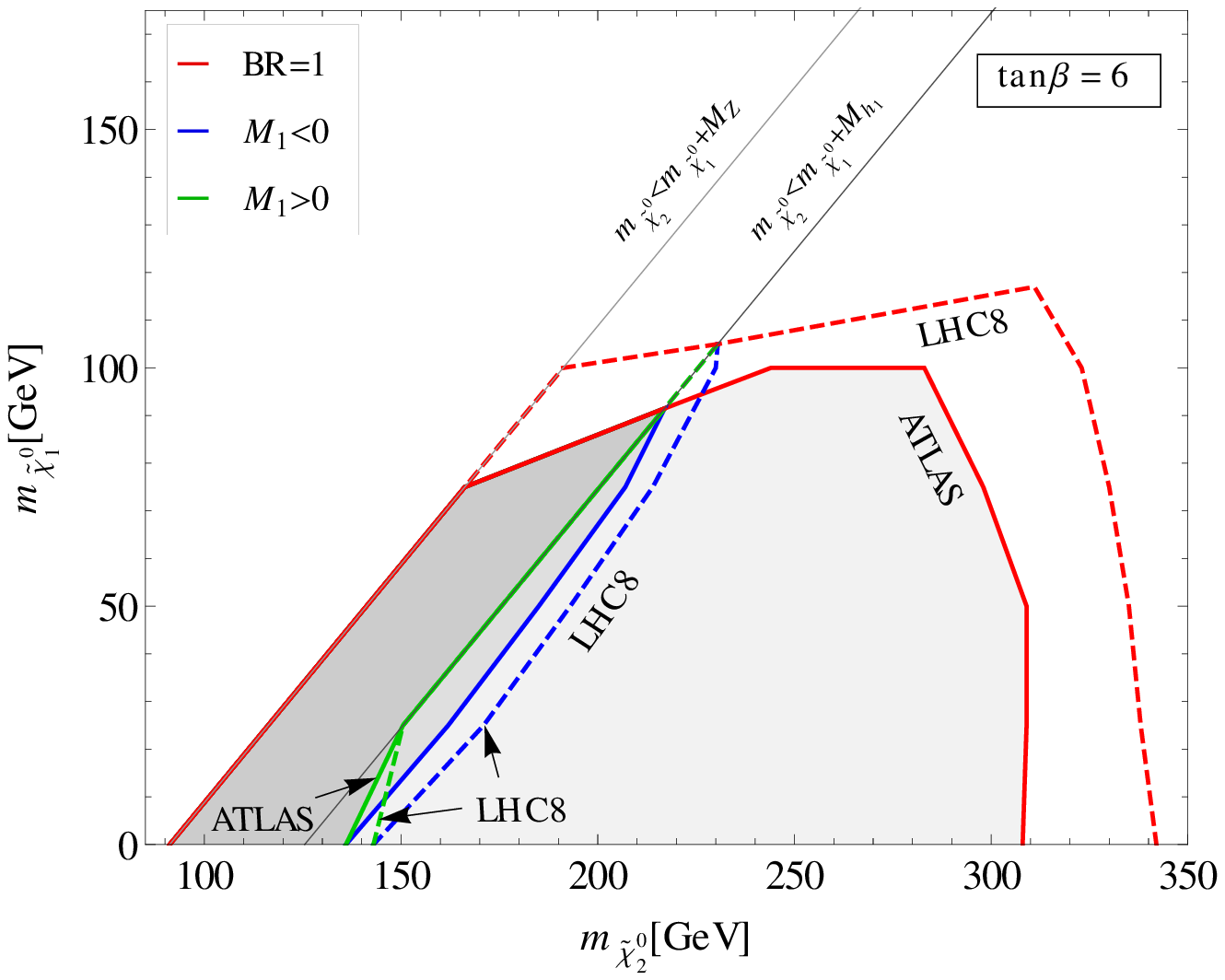}\\[1em]
\includegraphics[width=0.70\textwidth]{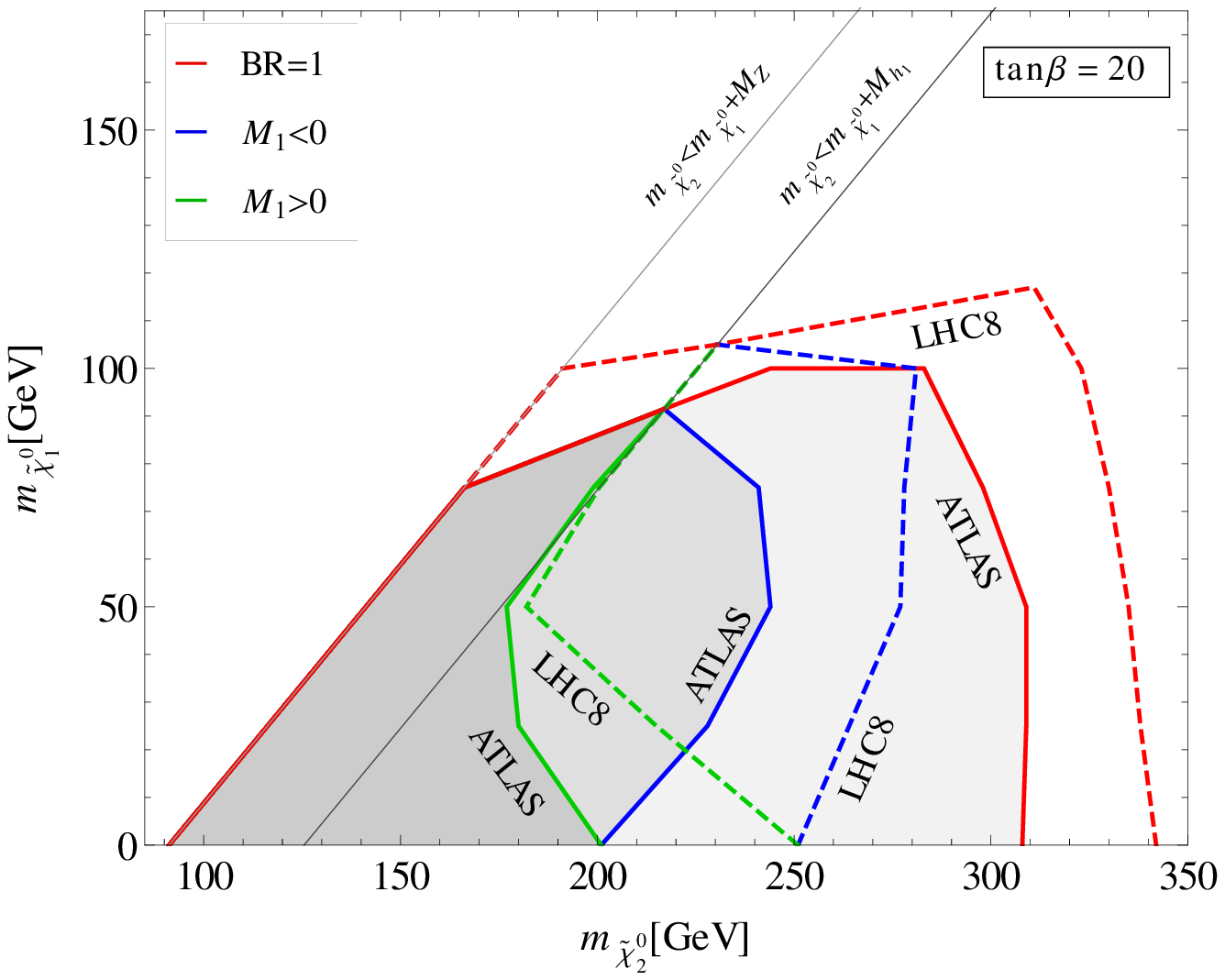}
\vspace{-1em}
\caption{
Contours showing the approximate excluded region from \Satlas\
in the $\mneu{2}$--$\mneu{1}$ plane
with $\tb = 6$ (upper plot) and  $\tb = 20$ (lower plot).
The solid lines (shaded areas)
correspond to the exclusion for the analyzed ATLAS
data~\cite{ATLAS:2012-154}, and 
the dashed lines indicate our projection for the combined LHC8 data,
both for the case where it is assumed $\br(\neu{2}\to\neu{1}Z)=1$,  
and where the decays $\neu{2} \to \neu{1} \He$ are taken into account
for $\MOne > 0$ (green), and for $\MOne < 0$ (blue) as indicated,
calculating $\br(\neu{2}\to\neu{1}Z)$ at NLO.  
Above the light (dark) gray line the on-shell
decay $\neu{2} \to \neu{1} Z (\He)$ is kinematically forbidden.
}
\label{fig:contour-LHC8}
\end{center}
\vspace{-2em}
\end{figure}

The main results are summarized in \reffi{fig:contour-LHC8}, 
where the exclusion region is shown in the $\mneu{2}$--$\mneu{1}$ plane
for $\tb=6$ (upper plot) and $\tb=20$ (lower plot). 
The solid lines (shaded areas) correspond to the data presented in
\citere{ATLAS:2012-154}.  
The dashed lines are the projection for the combination of ATLAS
and CMS LHC8 data, calculated as described in \citere{Bharucha:2013epa}.
The red lines show the ATLAS analysis, the green lines take
into account the decays $\neu{2} \to \neu{1} \He$ for $\MOne > 0$, and
the blue ones for $\MOne < 0$. 
The exclusion curves are not smooth, reflecting the fact that excluded
cross sections obtained from ATLAS are only available for a sparse grid of
points in the $\mneu{2}$--$\mneu{1}$ plane, and are given both for the
ATLAS data and for LHC8 combined data. 
Technically, this is achieved by interpolating the
cross section as a function of $\mneu{1}$ for fixed values of $\mneu{2}$.
Note that above the light (dark) gray line the on-shell
decay $\neu{2} \to \neu{1} Z (\He)$ is kinematically forbidden.

The results for 
scenario \Satlas, i.e.\ with $\tb=6$, are shown in the upper figure. 
The area excluded by ATLAS is drastically reduced when the decay 
$\neu{2} \to \neu{1} \He$ is taken into
account. Only the region where 
$\neu{2} \to \neu{1} \He$ is kinematically forbidden, and a small
strip close to the kinematic limit can be excluded by the
current ATLAS analysis. The excluded area grows marginally taking
into account the projection for the LHC8 full data set, i.e.\ the 
projected combination of ATLAS and CMS data.

We show the results for scenario \Stb, i.e.\ with $\tb=20$, in the 
lower figure of \reffi{fig:contour-LHC8}.
While below the Higgs threshold the exclusions are independent of $\tb$,
above this threshold the excluded regions are
somewhat larger for $\tb = 20$. However, the excluded region is clearly
reduced in comparison to 
the simplified model case where the Higgs channel is neglected.
Again, \reffi{fig:contour-LHC8} can easily be understood via
\refeqs{eq:nnz.approx} and (\ref{eq:nnh.approx}), where we see that for
smaller $\tb$ and  
large $\mu$ the Higgs channel dominates, and the branching ratio to the
$Z$ boson is considerably smaller than one.

\begin{figure}[t!]
\begin{center}
\begin{tabular}{c}

\includegraphics[width=0.48\textwidth]{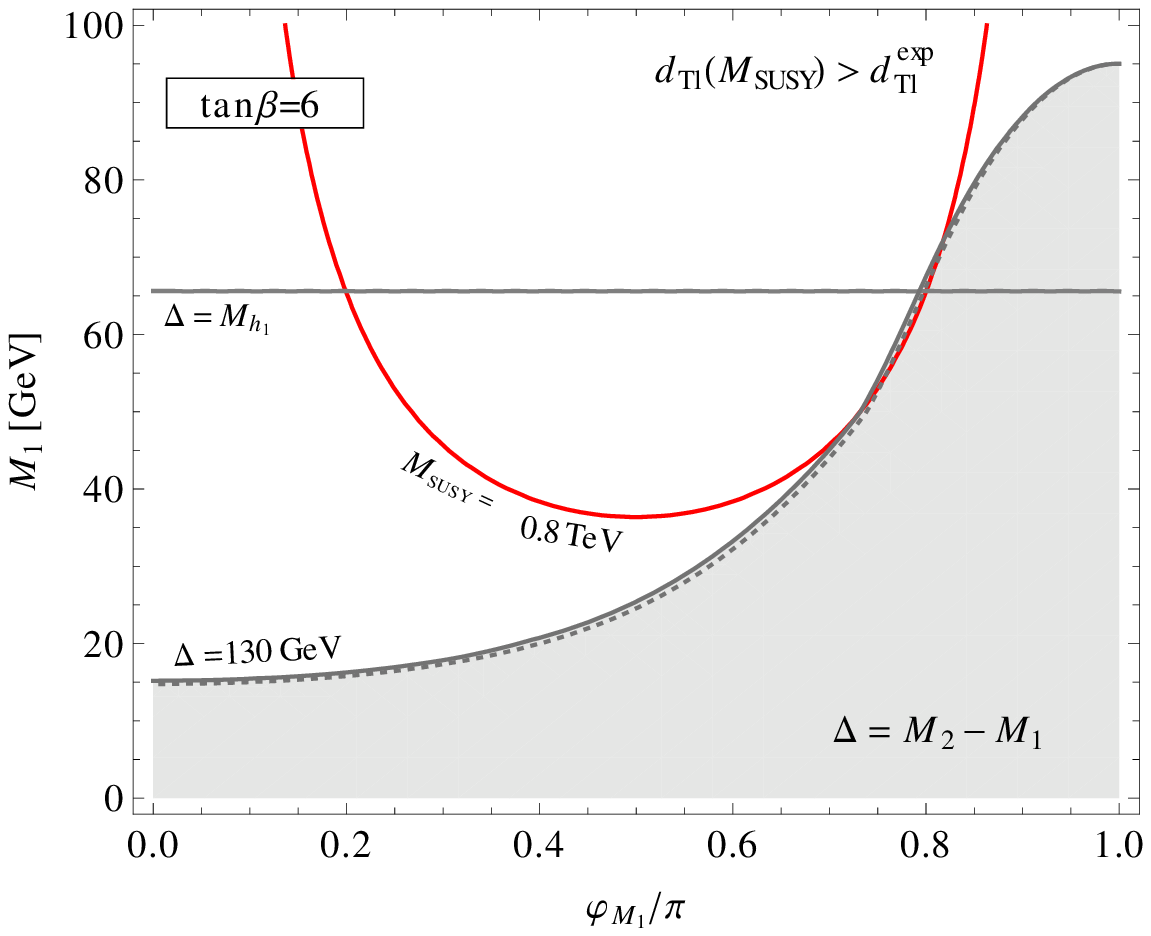}\hspace{.3cm}
\includegraphics[width=0.48\textwidth]{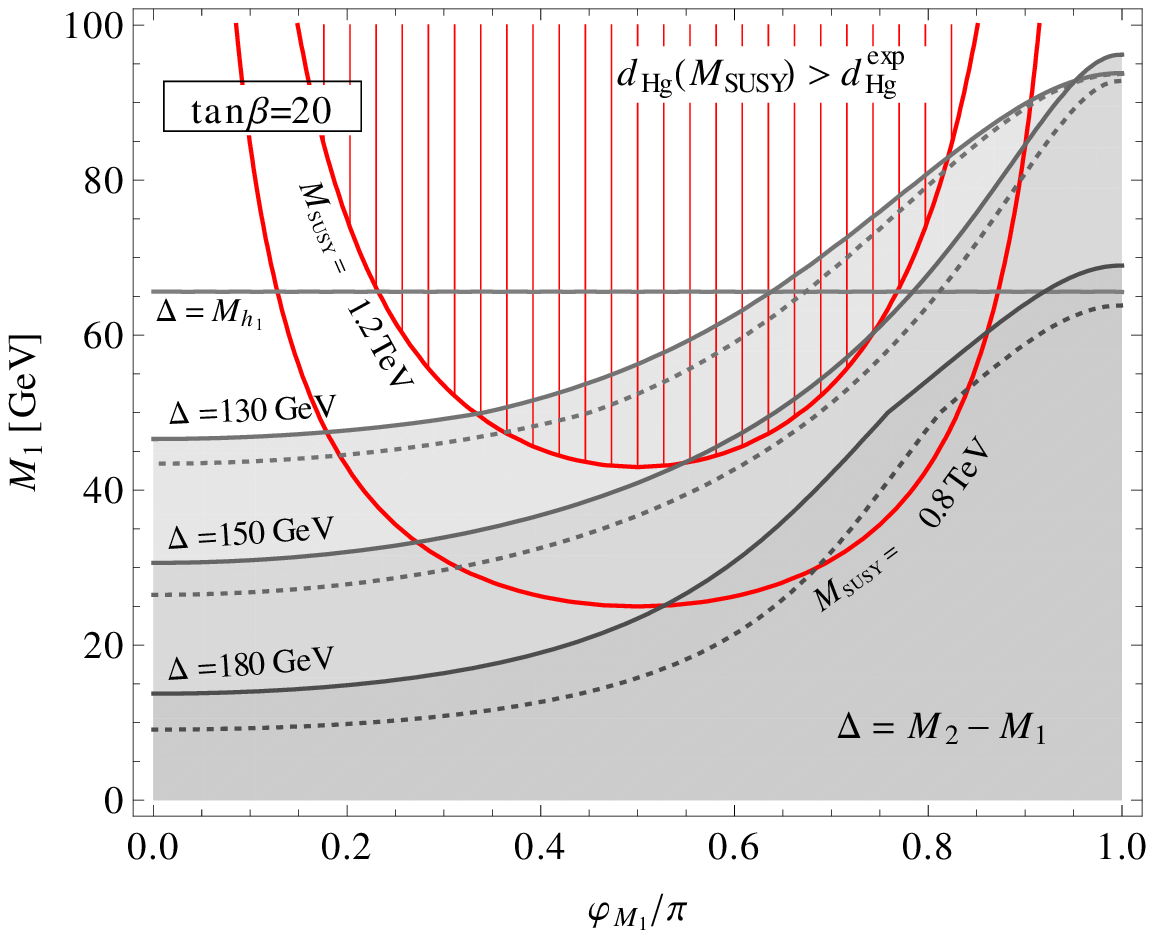}

\end{tabular}
\caption{
Contours showing the excluded region from currently analyzed ATLAS data for
\Satlas\ in the $\MOne$--$\phiMe$ plane, with $\tb = 6$ (left) and 
$\tb = 20$ (right). $\MTwo$ is fixed via $\De = \MTwo - \MOne$, which
corresponds approximately to  
diagonal lines in e.g.\ \reffi{fig:contour-LHC8}, parallel to the Higgs
threshold given by $\De = \MHe$. 
The solid (dotted) lines indicate that the exclusion
contours are calculated using NLO (tree-level) branching ratios for 
the $\neu2$ decays.
At $\De = 150 \gev$ for $\tb = 6$ and $\De = 210\gev$ for $\tb = 20$
there is no exclusion from ATLAS. 
The red lines define exclusion contours for \Ssusy, where $\msusy$ is
indicated (see text), from the EDMs of thallium ($d_{\rm Tl}$) and mercury
($d_{\rm Hg}$).}
\label{fig:contour-phiMe}
\end{center}
\end{figure}

As discussed in \refse{sec:notation}, the neutralino-Higgs coupling
decreases with $\phiMe$, 
and in \refeq{eq:nnh.approx} we see the
sensitivity to the phase increases with $\tb$. 
This $\phiMe$ dependence affects the exclusion bounds on $\MOne$ and $\MTwo$,
as illustrated in \reffi{fig:contour-phiMe}, where we consider the 
ATLAS exclusion limits in the $\phiMe$--$\MOne$ plane, for $\tb = 6$
(left) and $ 20$ (right),
and for $\De := \MTwo - \MOne = \MHe,\,130,\,150,\,180 \gev$
(which defines the value of $\MTwo$ in the plots).
The values of $\Delta$ here correspond approximately to diagonal lines in the
$\mneu{2}$--$\mneu{1}$ plane in \reffi{fig:contour-LHC8}, starting from the
Higgs threshold at $\Delta=M_{h_1}$.
The solid (dotted) lines indicate the calculation is at NLO (tree-level).
The limits in red are given by the requirement that the EDMs 
for thallium and mercury, $d_{\rm Tl}(\msusy)$ and $d_{\rm Hg}(\msusy)$,
calculated using
\texttt{CPsuperH~2.3}~\cite{Lee:2012wa,Lee:2007gn,Lee:2003nta}%
\footnote{
Similar results can be obtained with
\texttt{FeynHiggs}~\cite{feynhiggs,mhiggslong,mhiggsAEC,mhcMSSMlong}.}%
~for a specific value of $\msusy$ within \Ssusy, are below the upper
limit, i.e.\  
$d^{\rm exp}_{\rm Tl} =9.0\,10^{-25}\, e\,\mathrm{cm}\,(90\%~\mathrm{CL})$ or 
$d^{\rm exp}_{\rm Hg} =3.1\,10^{-29}\, e\,\mathrm{cm}\,(95\%~\mathrm{CL})$~\cite{Regan:2002ta,Griffith:2009zz}.
We adopt a common mass scale $\msusy = \Msqez = \Msqd = \Msl$, although
the EDMs depend mainly on $\Msqez$ and $\Mslez$. 
We display the limit for the EDM that provides the
strongest bound, i.e.\ from $d_{\rm Tl}$ for $\tb=6$ and from
$d_{\rm Hg}$ for $\tb=20$. Although 
$\msusy=0.8\tev$ is disfavored at the LHC, we show these limits 
for comparison, as there is no exclusion 
from the EDMs for higher values of $\msusy$ (i.e. in \Satlas) for $\tb=6$.

From \reffi{fig:contour-phiMe} one can clearly see how
$\phiMe$ affects the exclusion limit on $M_1$, 
which is much stronger for $\phiMe=\pi$ than for $\phiMe=0$, as also seen in
\reffi{fig:contour-LHC8}. Furthermore, as discussed above, the
effect of the phase is clearly much more pronounced at lower values of $\tb$.
On varying $\phiMe$ from 0 to $\pi$, for $\De =130\gev$, the limit on $\MOne$ 
changes by $\sim 80\gev$ for $\tb=6$ in contrast to $50\gev$ for $\tb=20$.
Note that the exclusion disappears completely for $\De >135\gev$ for
$\tb=6$ and $\De >200 \gev$ for $\tb=20$. The $\De = \MHe$ line is also
shown in order to illustrate that below the Higgs threshold, 
the dependence on $\phiMe$ vanishes. 

From the right plot of \reffi{fig:contour-phiMe} we notice that the
one-loop corrections, i.e.\ the difference 
between solid and dotted lines, are clearly sizeable for $\tb = 20$,
shifting the excluded value of $\MOne$ (for fixed $\phiMe$)  by up to 
$\sim 12 \gev$ for $\De = 180 \gev$.


\section{Conclusions}

We have reviewed the exclusion limits on the parameters describing the
electroweak sector of the MSSM from direct $\cha1 \neu2$ production
searches via $WZ+E_T^{\rm miss}$ at the LHC. 
We started by considering the baseline scenario used by ATLAS in their
public note 
for $21~\ifb$~\cite{ATLAS:2013-035}, where it is assumed that the 
$\cha1$ and $\neu2$ decay 100\% via $\cha1 \to \neu1 W^\pm$ and 
$\neu2 \to \neu1 Z$, and limits on $\mneu2$ of up to $\sim 300 \gev$ are
derived. 
We investigated how these limits are affected when using NLO results
both for the SUSY production cross sections as well as for the branching
ratio calculations. 
We found that, apart from the region where
$\neu2 \to \neu1 \He$ is kinematically forbidden%
\footnote{
It is interesting to note that in that region the most recent ATLAS
analysis~\cite{Aad:2014vma} found a $\sim 1.5\,\si$ excess of events.}%
, only a very small
strip in the $\mneu2$--$\mneu1$ plane can be excluded in this baseline
scenario. Allowing the gaugino mass parameter $\MOne$
to take negative values (corresponding to $\phiMe = \pi$),
slightly larger regions in the $\mneu2$--$\mneu1$ plane can be
excluded. Going from the baseline value $\tb = 6$ to $\tb = 20$ again
leads to somewhat larger excluded regions, but the decay
$\neu2 \to \neu1 \He$ is still clearly seen to have a substantial effect on
the limits.  
We furthermore reviewed the dependence of the excluded mass
regions on the phase of $\MOne$, and found a strong dependence on $\phiMe$.
In the future, limits on $WZ+E_T^{\rm miss}$ and $Wh+E_T^{\rm miss}$ 
could also be exploited as a method to constrain $\phiMe$,
complementary to the EDMs.

\medskip
Altogether these results show, on the one hand, how important it is to
look at a realistic spectrum (i.e.\ where the decays to a Higgs boson
are not neglected), and on the other hand that dedicated searches for
the $Wh+E^{\rm miss}_T$ channel are beneficial~\cite{Baer:2012ts}.
The results indicate that there is ample room for chargino/neutralino
production at the ILC with $\sqrt{s} \le 1 \tev$.


\subsection*{Acknowledgments}
A.B.\ gratefully acknowledges support of the DFG through the grant SFB 676,
``Particles, Strings, and the Early Universe''. 
The work of S.H.\ was partially supported by CICYT (grant FPA
2010--22163-C02-01). 
S.H.\ and F.v.d.P.\ were supported by 
the Spanish MICINN's Consolider-Ingenio 2010 Programme under grant
MultiDark CSD2009-00064. 
We thank the GRID computing network at IFCA for 
technical help with the OpenStack cloud infrastructure~\cite{Campos:2012vb}.


\end{document}